\documentclass[10pt]{iopart}
\usepackage{iopams}  

\usepackage{subfig}
\usepackage{graphicx}
\usepackage{xcolor}
\usepackage{siunitx}
\usepackage{cite}

\newcommand{\bra}[1]{\left\langle #1\right|}
\newcommand{\ket}[1]{\left| #1\right\rangle}
\newcommand{\nangle}[1]{\left\langle #1 \right\rangle}

\usepackage[normalem]{ulem} % for sout{} command

\begin{document}

\title[Electron Extraction with Higher Order Coulomb Scattering]
{Electron-Extraction from Excited Quantum Dots with Higher Order Coulomb Scattering}

\author{Alex Arash Sand Kalaee and Andreas Wacker}

\address{Division of Mathematical Physics, Lund University, Lund, Sweden}
\ead{kalaee@teorfys.lu.se}
\vspace{10pt}
\begin{indented}
\item[]\today
\end{indented}

%%%%%%%%%%%%%%%%%%%%%%%%%%%%%%%%%%%%%%%%%%%%%%%%
%
%	ABSTRACT
%
%%%%%%%%%%%%%%%%%%%%%%%%%%%%%%%%%%%%%%%%%%%%%%%%

\begin{abstract}
The electron kinetics in nanowire-based hot-carrier solar cells is
studied, where both relaxation and extraction are considered concurrently.
Our kinetics is formulated in the many-particle basis of the interacting system. 
Detailed comparison with simplified calculations based on product states shows 
that this includes the Coulomb interaction both in lowest and higher orders.
While relaxation rates of 1 ps are obtained, if lowest order processes are 
available, timescales of tens of ps arise if these are not allowed for
particular designs and initial conditions.
Based on these calculations we {quantify the second order effects and}
discuss the extraction efficiency,
which remains low unless an energy filter by resonant tunnelling is applied.
\end{abstract}

\vspace{2pc}
\submitto{\NT}
%\maketitle
\ioptwocol

%%%%%%%%%%%%%%%%%%%%%%%%%%%%%%%%%%%%%%%%%%%%%%%%
%
%	INTRODUCTION
%
%%%%%%%%%%%%%%%%%%%%%%%%%%%%%%%%%%%%%%%%%%%%%%%%
\section{Introduction}
The solar spectrum covers a wide range of energies. In contrast, for an
optoelectronic device, we only achieve a good efficiency by adding electrons
or holes at the respective electrochemical potential of the contacts
\cite{WurfelJPhotovolt2015,LimpertNJP2015}.
Thus, only a defined exciton energy is efficiently converted to electrical energy.
This is the origin of the \emph{ultimate efficiency limit} of
Shockley and Queisser \cite{ShockleyJAP1961}.

The idea both behind hot-carrier solar cells
\cite{RossJAP1982,GreenProgPhotovolt2001,WurfelProgPhotovolt2005} or 
{Multiple Exciton Generation (MEG)} devices \cite{KolodinskiAPL1993,LandsbergJAP1993,NozikPhysicaE2002,GoodwinNanoph2017,YanNatEn2017} is to combine
different excitonic energies to the optimal one by employing the inter-particle Coulomb interaction
and extract those at the
corresponding difference of electrochemical potentials.
The conversion efficiency
of hot-carrier solar cells has been extensively studied under different conditions
\cite{AlibertiJAP2010,TakedaJAP2009,FengAPL2012,LimpertNJP2015,KahmannJMCC2019}.
A big contribution to conversion loss in a hot-carrier solar cell is the
thermalisation between the hot electron and the lattice, an objective is thus
to increase the thermalisation time in the cell
\cite{TakedaAPE2010,LeBrisAPL2010,LuqueSEMSC2010,TsaiProgPhotovolt2018}.
Further, it has been argued that the hot-carrier cooling rate can be slowed down
by a combination of phonon bottleneck \cite{BockelmannPRB1990} and reduced accessibility of the phase
space by employing nanostructures like quantum wells or quantum dots (QDs)
rather than their bulk counterpart
\cite{PelouchPRB1992,RosenwaksPRB1993,ConibeerSEMSC2015,ConibeerJJAP2017,NguyenNatEn2018}.
Note that in addition to the extended hot-carrier cooling time, nanowire
heterostructure solar cells also exhibit tunable photoabsorption beyond that
of the bulk analogue
\cite{CaoNatMat2009,CaoNL2010,WuNL2012,KimNL2012,SvenssonNL2013,BrongersmaNatMat2014}.

A source of the slowed cooling rate in low dimensional nanostructures is the
quantization effect from confinement of the electronic wave function,
which affects the possibility of Auger processes between the carriers
\cite{NozikPhysicaE2002,BeardNL2007}.
The Auger process is in turn mediated by the Coulomb interaction which
must be included when modelling carrier dynamics.
Studies on carrier dynamics after MEG have employed master equations
which include the Coulomb interaction to first order
\cite{ShabaevNL2006,FranceschettiNL2006}.
Here we seek to describe the solar cell using a master equation
approach that includes the Coulomb interaction to higher orders.

In order to efficiently employ these hot-carrier schemes energetically narrow
filters must be implemented which allow for selective extraction of excitons
\cite{FengAPL2012,WurfelSEMSC1997,AlibertiAPL2011,TakedaJAP2015}.
The hot-carrier solar cell thus requires an
appropriate kinetics and properly matched extraction
processes, which we study in detail in this article.
Here we only concentrate on the electronic part, assuming that the holes are
generated at the top of the valence band and are collected from there.

Following the discussion above, we divide our study in three parts: At
first we consider the kinetics of electrons after photoexcitation with a high
photon energy, where we study the rate of formation of multiple excited
electrons after an initial high energy excitation.
This process is dominated by the electron-electron interaction and we show that
both first-order scattering processes and higher-order contributions play an
important role. At second we discuss the features of the extraction kinetics
through a single barrier and show that for a system dominated by second order
transitions the extraction kinetics is impaired compared to first order.
Finally we show how to amend the impairment by implementing a narrow energy
filter, i.e. changing the barrier geometry to a double barrier potential.

%%%%%%%%%%%%%%%%%%%%%%%%%%%%%%%%%%%%%%%%%%%%%%%%
%
%	METHODS
%
%%%%%%%%%%%%%%%%%%%%%%%%%%%%%%%%%%%%%%%%%%%%%%%%
\section{Methods and Model}
To study the extraction kinetics in the nanowire QD we need to
time-evolve the electronic quantum states in the presence of
Coulomb  interactions, relaxation processes, and
tunnelling to electron reservoir regions. For this purpose we
first determine the single{-}particle states of the
electrons in the QD and evaluate the Coulomb interaction between these.
For the electronic eigenstates of the interacting system we consider
time-evolution of density matrix within the Lindblad master equation
\cite{LindbladCMP1976}, where dephasing and extraction is treated
via the PERLind approach \cite{KirsanskasPRB2018}.

\subsection{Geometry and Single{-}Particle Levels}
\label{sec:meth:geom}
\begin{figure*}
\centering
\includegraphics[width=0.8\linewidth]{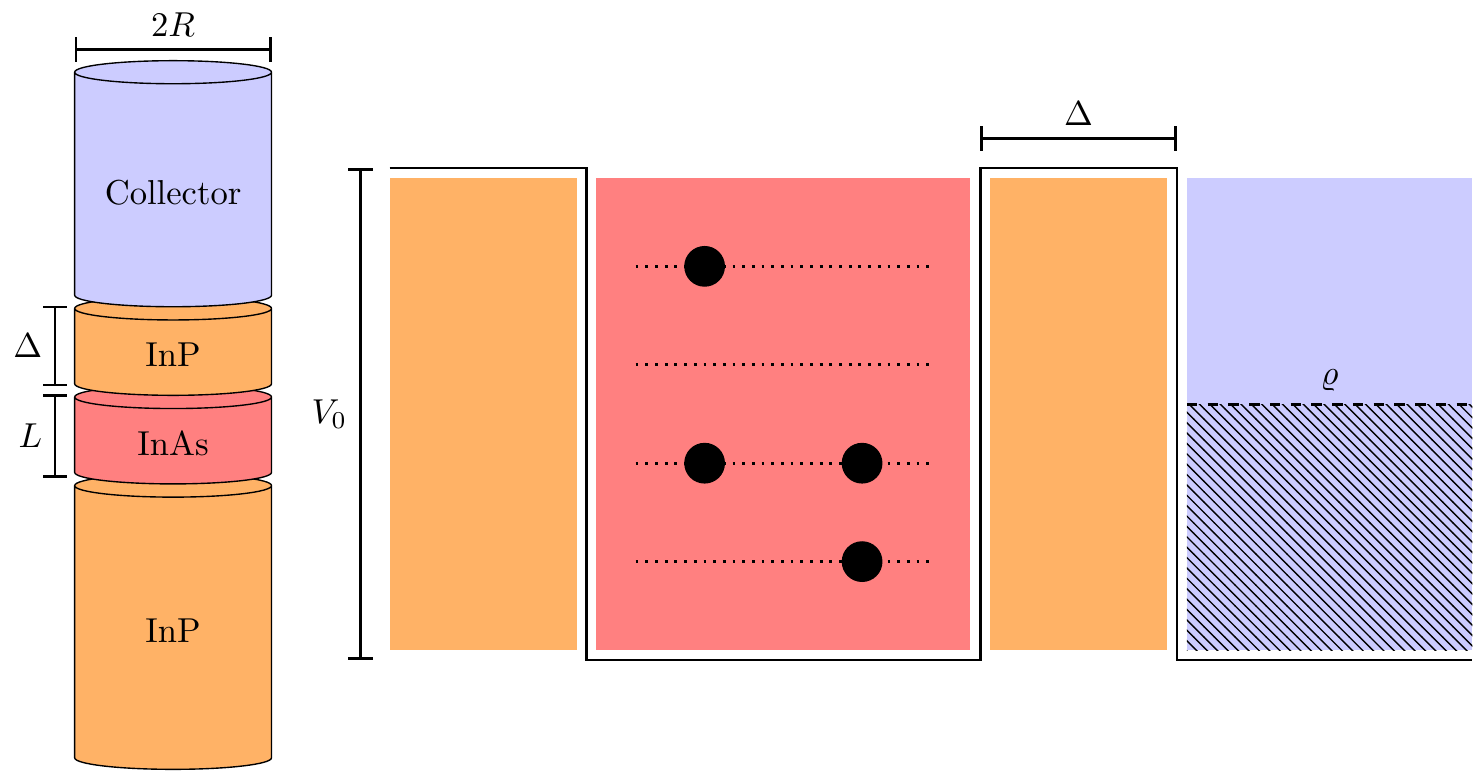}
\caption{Geometry (left) and energy (right) schematics of the nanowire QD system. The QD is
assumed to be a cylindrical piece of InAs with height $L$ and radius $R$.
To the one side it is connected to a collecting contact region
after an InP barrier of length $\Delta$.}
\label{fig:schema}
\end{figure*}

The InAs/InP nanowire axial heterostructure constitutes an interesting
system for the solar energy conversion in nanostructures\cite{LimpertNL2017}.
We model the QD as an InAs cylinder with height  $L=\SI{8}{\nano\meter}$ and
radius $R=\SI{8}{\nano\meter}$  between InP barriers forming
a finite  well in the conduction band, see Figure~\ref{fig:schema}.
{A similar system has been realised and shown to function as a hot-carrier solar cell \cite{LimpertNL2017,LimpertNT2017}.}
While the dimensions are chosen slightly smaller than typical structures, 
this yields a model system for which the calculation of the many-particle
eigen-states is tractable. 
{Likewise, we restrict our model to include only the electrons in the conduction band,
and thus exclude valence band holes, for the sake of computational complexity.}
We assume that the conduction band discontinuity
between InAs and InP in a thin nanowire provides a potential well of depth
$V_0 =\SI{0.7}{\e\volt}$, which is of the same order of magnitude as the
observed discontinuity
\cite{ZervosJAP2004,BjorkAPL2002a,BjorkAPL2002b}.
Finally, the effective masses are $m_\mathrm{InAs} = 0.023 m_e$
\cite{NakwaskiPhysicaB1995} and
$m_\mathrm{InP} = 0.073m_e$ \cite{MossPhysica1954},
where $m_e$ is the free electron mass.

To consider extraction we connect one side of the QD to a collecting contact
region with potential offset $\varrho$ which prohibits electrons with
energy lower than this level to be extracted from the QD,
see Figure~\ref{fig:schema}. This can be achieved using a ternary alloy.
{We model this region as a semiconductor with a
  conduction band minimum at $\varrho$ and apply the effective mass of InAs, for simplicity. }

\begin{figure*}
\centering
\includegraphics[width=0.8\linewidth]{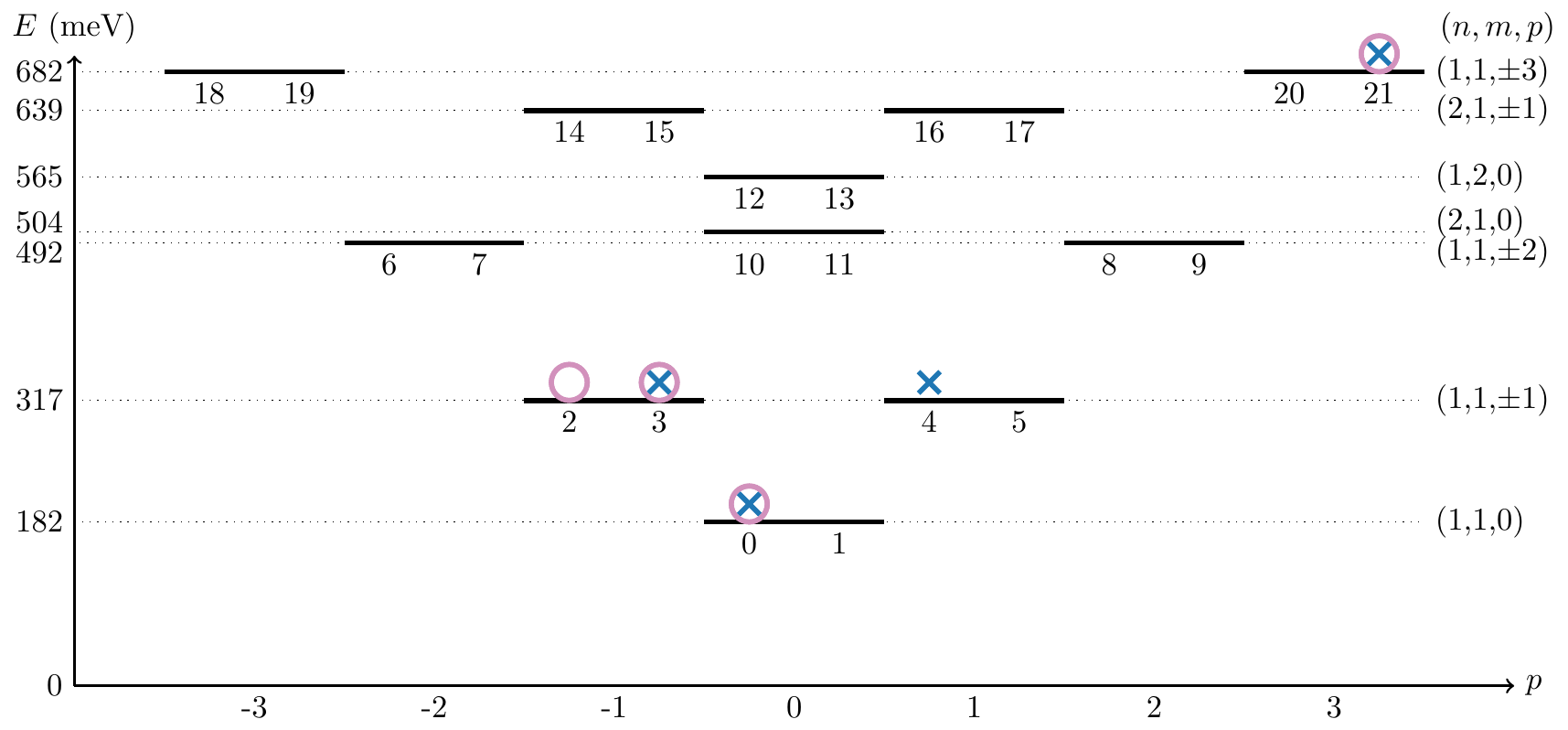}
\caption{Single{-}particle levels for the cylindrical InAs QD with
height $L=\SI{8}{\nano\meter}$ and radius $R=\SI{8}{\nano\meter}$.
Every solid line represents a confinement orbital, even numbered states have
spin up and odd spin down. The quantum numbers $(n,m,p)$ describe the axial,
radial and angular quantization respectively. The blue crosses and violet circles
represent the initial states $\ket{A}=\ket{0,3,4,21}$ and
$\ket{B}=\ket{0,2,3,21}$ respectively, which we will consider in Sec.~\ref{SecResults}.}
\label{fig:states}
\end{figure*}

The single{-}particle levels provided in Figure~\ref{fig:states}
are evaluated
using heterostructure envelope functions in the effective mass approximation
\cite{BenDanielPR1966}.
{
Integrating the density of the electronic envelope function over the volume contained within a shell of \SI{1}{nm} thickness around the quantum dot we obtain norms above 0.95 for all states --
more than 95 percent of the electronic wavefunction is contained within the shell.
}
This allows us to consider geometries with barriers
as thin as $\Delta = \SI{1}{\nano\meter}$ while still applying a basis of
confinement states calculated for infinitely thick barriers.
\subsection{Coulomb Matrix Elements}
An important part of the electron kinetics in the QD stems from the
Coulomb electron-electron interaction, which strongly depends on the
spatial overlap of the envelope functions.
The general Coulomb matrix is defined in the occupation number representation as
\begin{eqnarray}
\hat{H}_{ee} &= \sum_{m < n,l<k} V^\mathrm{eff}_{mnkl}
\hat{a}^\dagger_m\hat{a}^\dagger_n\hat{a}_k\hat{a}_l,\\
V_{mnkl}^\mathrm{eff} &= V_{mnkl}-V_{mnlk}
\end{eqnarray}
{ Here the} $\hat{a}_i$ are the annihilation operators for electrons in state
$i$ and the Coulomb elements are defined
% cannot use \iint in iopart, will resort to \int\!\!\!\int
\begin{eqnarray}
\fl V_{mnkl} =\nangle{\chi_m|\chi_l}\nangle{\chi_n|\chi_k}
\frac{e^2}{4\pi\varepsilon\varepsilon_0}\nonumber\\
\times\int\!\!\!\int \rmd^3\mathbf r d^3\mathbf r'
\varphi^*_m(\mathbf r) \varphi_n^*(\mathbf r')
\frac{1}{|\mathbf r - \mathbf r'|}
\varphi_k(\mathbf r') \varphi_l(\mathbf r)
\end{eqnarray}
where $\ket{\chi}$ is the electron's spin state and $\varphi$ the spatial
envelope wave function, $\varepsilon_0$ is the vacuum permittivity of free
space and $\varepsilon$ the relative permittivity of the material.
As the bulk of the electron density is contained within the InAs structure,
we use the relative permittivity $\varepsilon_\mathrm{InAs}=15$.
Further, the considered orbitals all belong to the same band, hence, we can
neglect effects resulting from the lattice periodic wave function at the
atomic scale.
The Coulomb elements are obtained by Monte Carlo integration with importance
sampling under the VEGAS routine \cite{GalassiBook2018}.

Important Coulomb elements are the direct elements and orbital exchange
elements, typical values for the two are, respectively,
\begin{equation}
V_{1,3,3,1} = \SI{15}{meV}, \qquad V_{1,3,1,3} = {\SI{3.8}{meV}}.
\end{equation}
The size of the direct element can be understood as the charging energy
$e^2/C$ of a spherical capacitor with capacitance
$C=4\pi\epsilon_0\varepsilon_\mathrm{InAs}r_\mathrm{eff}$. Here the
energy of \SI{15}{\milli\e\volt} corresponds to the
effective radius  $r_\mathrm{eff}=\SI{6.2}{\nano\meter}$.
The value of $r_\mathrm{eff}$  agrees  well with the root mean square
distances of the occupation probabilities
from the centre of the cylinder, \SI{5.2}{\nano\meter} for level 1
and \SI{6.3}{\nano\meter} for level 3.
The exchange element corresponds to a reduction by a factor of 4 from its
respective direct element; this is a typical difference in scale between
pairs of direct and exchange elements.

\subsection{Density Operator Master Equation}
Having determined the electronic single{-}particle states of the QD we are
now in a position to construct the system Hamiltonian and density matrix
which we use to perform a Lindbladian time evolution of the system.

From the single{-}particle orbital energies and the Coulomb matrix we can
assemble the full system Hamiltonian
\begin{equation}
\hat{H} = \sum_i E_i \hat{a}^\dagger_i\hat{a}_i + \hat{H}_{ee}.
\end{equation}
The state of the system is described by the reduced density operator
$\hat{\rho}_S$. When considering the density operator we will neglect
coherences between different particle numbers,
the operator can thus be further reduced to a tensor product,
\begin{equation}
\hat{\rho}_S = \hat{\rho}_0\otimes\hat{\rho}_1\otimes\ldots\otimes
\hat{\rho}_N,
\end{equation}
where $\hat{\rho}_n$ is the reduced density operator for $n$ particles and $N$
is the highest number of particles in the system.

Following the method used in Refs.~\cite{DamtieJPhysConfSer2016,DamtieJChemPhys2016}
we model the time evolution of the density operator during the extraction of
electrons using the Lindblad master equation 
\begin{equation}
\hbar\frac{\rmd}{\rmd t}\hat{\rho}_S = i[\hat{\rho}_S,\hat{H}] +
\sum_j^{N_\mathrm{jump}}\Gamma_j
\left[\hat{L}_j\hat{\rho}_S\hat{L}_j^\dagger -
\frac{1}{2}\left\{\hat{\rho}_S,\hat{L}_j^\dagger\hat{L}_j\right\}\right]
\label{EqLindblad}
\end{equation}
where $N_\mathrm{jump}$ is the number of jump operators $\hat{L}_j$ with rates
$\Gamma_j/\hbar$.

We define the Lindblad jump operators phenomenologically, where we
consider spin-conserving dephasing operators on each orbital $\alpha$,
\begin{equation}
\hat{L}^\alpha_\mathrm{deph} =
\hat{a}_{2\alpha}^\dagger\hat{a}_{2\alpha}+
\hat{a}_{2\alpha+1}^\dagger\hat{a}_{2\alpha+1},
\label{eq:meth:jump:deph}
\end{equation}
and the extraction operator on state $i$,
\begin{equation}
\hat{L}^i_\mathrm{extr} = \hat{a}_i.
\label{eq:meth:jump:extr}
\end{equation}
which corresponds to electron transfer to the collecting reservoir.

The dephasing strength, $\Gamma_\mathrm{deph} =\SI{6}{\milli\e\volt}$,
corresponds to a dephasing time of the order of \SI{100}{\femto\second}.
This is a typical time scale for these kinetic systems \cite{JasiakNJP2009}.
Considering the extraction strengths we interpret the electron as a classical
particle moving back and forth in the dot. The axial kinetic energy thus
provides the velocity of the particle in axial direction from which an
attempt rate is derived \cite{PriceAJP1998},
\begin{equation}
\nu_i = \frac{1}{L}\sqrt{\frac{E_{i,z}}{2m_\mathrm{InAs}}},
\end{equation}
where $E_{i,z}$ is the axial kinetic energy of the envelope state $i$.
For the two axial states, $n = 1,2$, the attempt rates are
\begin{equation}
\nu_{n=1}=\SI{75}{\pico\second^{-1}},\qquad
\nu_{n=2}=\SI{158}{\pico\second^{-1}}.
\end{equation}
The extraction strength, or \emph{transmission} strength,
is determined from the attempt rate as
\begin{equation}
\Gamma^i_\mathrm{extr} = \hbar\nu_i
\end{equation}
Note that these are not the final jump operators applied in this work.
We will further adjust these with the tunnelling probability in the following
section where the energetic information is taken into account.

\subsection{Position and Energy Resolving Lindblad Approach}
\label{sec:meth:perlind}
The Lindblad jump operators in \Eref{eq:meth:jump:deph}   and
\Eref{eq:meth:jump:extr} do not carry energetic
information about the interactions between the system and bath.
A scheme to incorporate these aspects is the Position and Energy Resolving
Lindblad approach (PERLind) \cite{KirsanskasPRB2018}.
Let the states $\ket{a},\ket{b},$ etc. be an eigenbasis of $\hat{H}$ with
$E_a,E_b,$ etc. as the corresponding eigenenergies. Then we
transform the matrix elements of every jump operator under the PERLind scheme
according to
\begin{equation}
\bra{a}\widetilde{L}_j\ket{b} = \sqrt{f_j(E_a - E_b)}\bra{a}\hat{L}_j\ket{b}.
\end{equation}
Where the function $f_j(E)$ describes energy dependence of the
jump transition $j$, where $E$ is the energy difference between the final and
the initial state. This defines new operators $\widetilde{L}_j$ which replace
the original operators $\hat{L}_j$ in the Lindblad master equation.

For the dephasing operator we assume the interaction to stem from coupling
to a phonon bath with temperature $T=\SI{300}{K}$. Here we use the
 heuristic  function
\begin{equation}
f_\mathrm{deph}(E) = \frac{E/k_BT}{\exp(E/k_BT)-1}\Theta(D_\mathrm{InP}-|E|),
\label{eq:meth:perl}
\end{equation}
which contains the Bose-distribution $f_\textrm{Bose}(E)$
for positive energies (absorption of phonons) and  $1+f_\textrm{Bose}(|E|)$ for negative $E$ (emission of phonons). The factor $|E|/k_BT$ reflects the phonon density of states and the energy dependence of the coupling. 
$f_\mathrm{deph}(E)$ is bounded by the Heaviside step function, $\Theta$,
to exclude phonons beyond the Debye energy for InP,
$D_\mathrm{InP}=\SI{37}{\milli\e\volt}$ \cite{AdachiBookChap2005}, which is approximately the largest phonon energy available.
Note that dephasing under this model can only connect eigenstates that are
separated in energy by less than $D_\mathrm{InP}$.

For modifying the extraction we model this process as tunnelling in a finite
heterostructure. The modifying function is
\begin{equation}
f_\mathrm{finite}(E)=\mathcal{T}(-E)\Theta(-E-\varrho)
\end{equation}
where $\mathcal{T}(E_e)$ is the transmission amplitude
described by Tsu and Esaki \cite{TsuAPL1973}. Note that the electron energy $E_e$ is taken from the  electronic system in the quantum dot so that its change in energy is $E=-E_e$ during the jump process)

%%%%%%%%%%%%%%%%%%%%%%%%%%%%%%%%%%%%%%%%%%%%%%%%
%
%	RESULTS
%
%%%%%%%%%%%%%%%%%%%%%%%%%%%%%%%%%%%%%%%%%%%%%%%%
\section{Results}\label{SecResults}
{
In order to obtain an understanding of the relaxation and extraction dynamics we consider qualitatively different initial states of the system with singly excited electrons.}
Specifically, we choose the product states
$\ket{A}=\ket{0,3,4,21}$ and $\ket{B}=\ket{0,2,3,21}$, which we assume
to be generated by some single{-}particle excitation process not specified
here.
{We consider these states to be two representative cases of the span of possible excited carrier states.}
State $\ket{A}$ is close in energy with
$\ket{0,3,8,9}$ and $V_{8,9,21,4}=\SI{4.4}{meV}$, which results in a
fast Auger process. For $\ket{B}$, the corresponding Coulomb matrix element
${V_{8,9,21,2}=0}$ due to conservation of angular momentum,
which should provide a longer lifetime. These states will be used to examine
thermalisation processes in the QD and subsequently to consider
the extraction kinetics. Here we will treat three different aspects
subsequently: (i) the quantum kinetics implied by the
electron-electron-interaction, (ii) the thermalisation induced  by our choice
of energy dependent dephasing, and (iii) the extraction of carriers.

\subsection{Quantum beating in the QD}
\label{SecBeating}
\begin{figure}
\centering
\subfloat{\includegraphics[width=\linewidth]{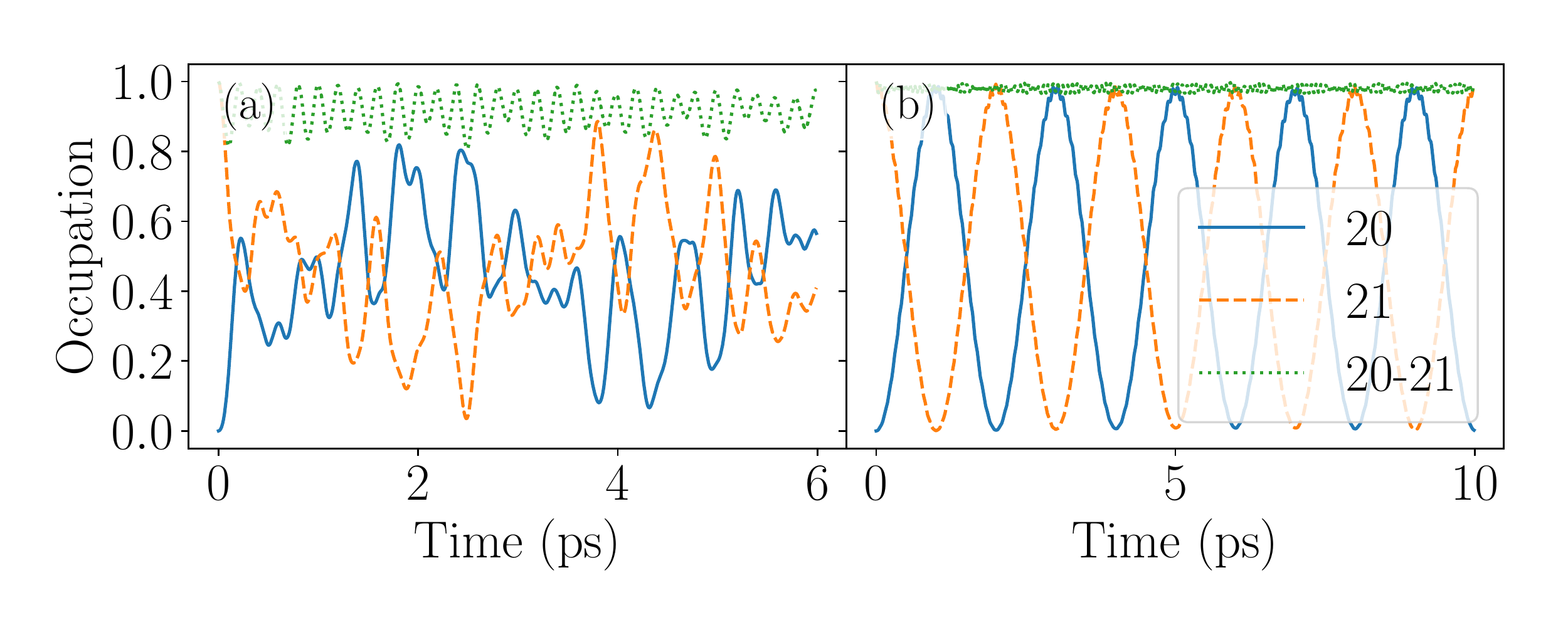}}\\
\vspace{-2em}
\subfloat{\includegraphics[width=\linewidth]{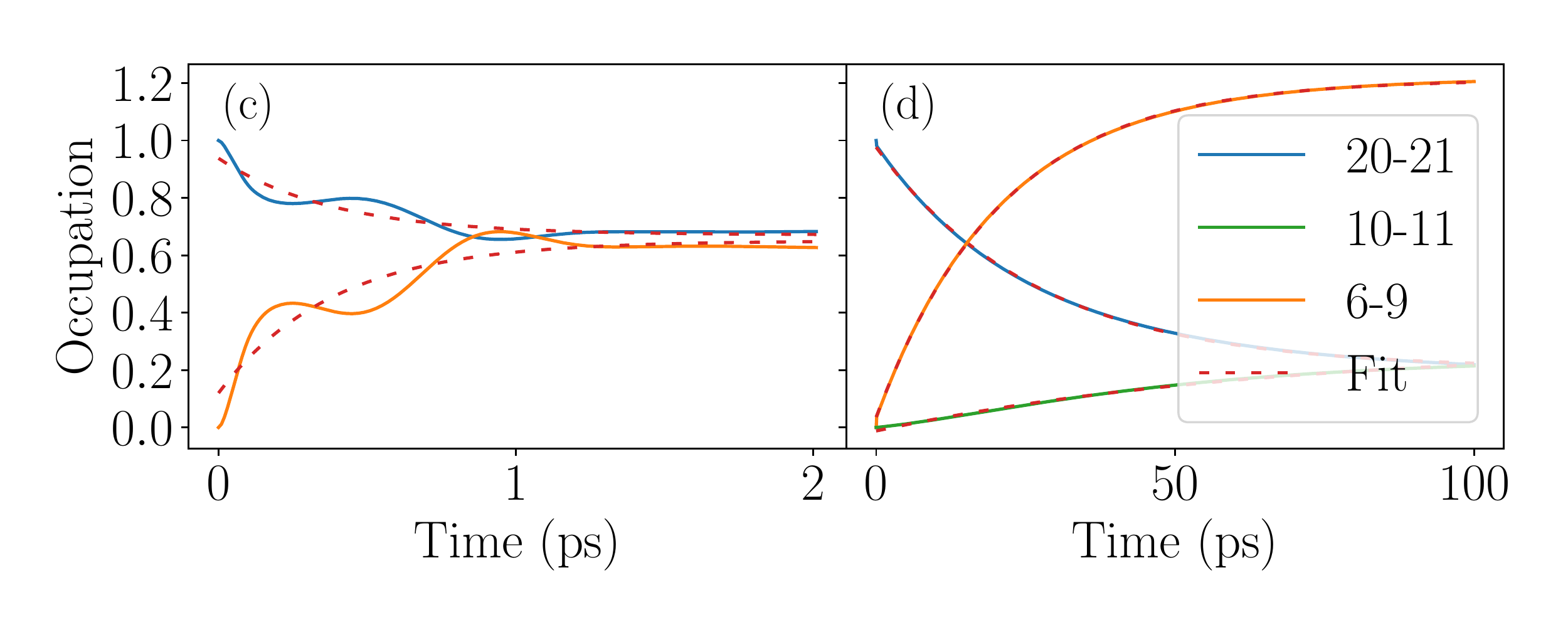}}
\caption{Time-dependence of  occupations of selected
single{-}particle  levels as denoted in Fig.~\ref{fig:states} for
the initial states $\ket{A}=\ket{0,3,4,21}$ (left) and $\ket{B}=\ket{0,2,3,21}$
(right).  In (a) and (b), dephasing and extraction is disregarded
and the pure quantum evolution shows distinct quantum beatings, which
mainly exchanges the populations of states 20 and 21 (as well as 1 and
0, not shown). Panels (c) and (d) show the corresponding results in
the presence of dephasing $\Gamma_\textrm{deph}=6$~meV, where we sum
the occupations over groups of states strongly coupled by the quantum
beating. Here the  dashed red lines provide exponential fits to the
respective occupations, where we extracted the thermalisation times
\SI{0.38}{ps} for draining of the hot-carrier levels 20-21 (blue)
and \SI{0.38}{ps} for filling of levels 6-9 (orange) during the
evolution of $\ket{A}$ in panel (c).  The alternative initial
state $\ket{B}$ in panel (d) provides \SI{26}{ps} for the
draining of hot-carrier levels 20-21, \SI{55}{ps} for filling of
levels 10-11 and \SI{20}{ps} for levels 6-9.}
\label{fig:therm}
\end{figure}

Figure~\ref{fig:therm}(a,b) shows the evolution of single{-}particle occupations
upon quantum evolution of \eref{EqLindblad} under the neglect of all
jump operators. Here the electron-electron interaction couples the states
with each other leading to quantum beating between different states with
similar energy. Most importantly, the pair of levels $(0,21)$, which is
occupied initially in both cases, is coupled to $(1,20)$ via the matrix element
\begin{equation}
V_{20,1,21,0}=\SI{0.781}{meV},
\end{equation}
This results in a beating between both occupations, as can be seen in
Figs.~\ref{fig:therm}(a,b) for both initial states. While the
occupations of levels 20 and 21 strongly alter in time, their sum
remains close to 1. For the initial state  $\ket{A}$ a complex
quasi-periodic scenario occurs, see Fig.~\ref{fig:therm}(a), due to a
second pair of equivalent levels  $(4,21)\leftrightarrow (20,5)$.
Additionally, further states are close to resonance here. In
contrast, Fig.~\ref{fig:therm}(b) displays an almost ideal beating
between the states $\ket{B}$ and $\ket{1,2,3,20}$ with the frequency
$f^B_\textrm{beat}=0.49$ THz. Surprisingly, $f^B_\textrm{beat}$ differs
from  $2V_{20,1,21,0}/h\approx 0.38$ THz, which we expect from the
direct interaction between these states. This difference can be
attributed to higher orders in the electron-electron interaction.
Actually, the observed frequency can be well reproduced by taking into
account the second order terms\cite{SakuraiBook1993}
in the effective matrix element between the beating states:
\begin{eqnarray}
&\bra{1,2,3,20}\hat{H}^{(2)}\ket{0,2,3,21}\nonumber\\
=&V_{20,1,21,0}+
\sum_{\ket{m}}\frac{\bra{1,2,3,20}\hat{H}\ket{m}\bra{m}\hat{H}\ket{0,2,3,21}}
{E_{\ket{0,2,3,21}}-E_{\ket{m}}}\nonumber\\
=& \SI{1.1}{meV}\approx \frac{h}{2} 0.51\textrm{ THz}\, ,
\label{EqFermi}
\end{eqnarray}
where the sum $\ket{m}$ runs over all different states which connect the initial
and final states. This demonstrates the high relevance of second-order
couplings for QD systems of typical sizes.

%%%%%%%%%%%%%%%%%%%%%%%%%%%%%%%%%%%%%%%%%%%%%%%%%%%%%%%%%%%%%%%
\subsection{Relaxation due to dephasing}
\label{sec:relax}
% First initial state
Upon introducing the dephasing jump operators in
\eref{EqLindblad}, we find relaxation processes to different
states as the coherent evolution is broken due to the coupling to
the environment.
In order to obtain signals relevant for extraction of electrons
we sum the occupations of levels, for which strong beating effects occur as
addressed in subsection~\ref{SecBeating}.
The results are shown in Fig.~\ref{fig:therm}(c,d) for the
initial states $\ket{A},\ket{B}$, respectively.

Assuming that the probability drift can be described as exponential
convergence, we are able to extract the thermalisation times of
the systems in Figure~\ref{fig:therm}(c,d).
For the initial state  $\ket{A}$, see, Figure~\ref{fig:therm}(c),
we find that the time scale of draining the hot-carrier levels 20-21
as well as the filling of levels 6-9 occur on the time scale
$\tau^A_\mathrm{drain}=\tau_\mathrm{fill}^A=$\SI{0.38}{ps}.
{These time scales are of the same magnitude as recently observed
  thermalisation and cooling times in highly excited InAs nanowires
  \cite{WittenbecherAbstract2018}.}
The scattering rate of a single Auger process with Coulomb element $V$ and detuning
$\Delta E$ is estimated using Fermi's golden rule
\begin{equation}
\Lambda =\frac{\left|V\right|^2}{\hbar} 2\pi\delta_{\widetilde\Gamma_\mathrm{deph}}(\Delta E)
\end{equation}
with Lorentzian broadening
\begin{equation}
2\pi\delta_{\widetilde\Gamma_\mathrm{deph}}(\Delta E) =
\frac{\widetilde\Gamma_\mathrm{deph}}{\Delta E^2+\widetilde\Gamma_\mathrm{deph}^2/4}.
\end{equation}
where we employ the dephasing strength
\begin{equation}
\widetilde\Gamma_\mathrm{deph} = \Gamma_\mathrm{deph}f_\mathrm{deph}(\Delta E).
\end{equation}
(The energy-dependence on $\Gamma_\mathrm{deph}$ is not strictly compatible with
the use of a Lorentzian. However, we refrain from a more elaborate description here.)

To estimate the effective scattering rates, we consider a subset of the Hilbert
space and have to take into account the multiplicities of the relevant
initial and final states, between which strong beatings occur as discussed above.
The initial state has to be treated as $N_i=6$ degenerate product states,
which are accessible by Coulomb matrix elements, while the final states
(with two of the levels 6-9 occupied) has $N_f=2$.
Furthermore, as two of the initial product states do not connect to the final states
there are $N_p=4$ possible Auger processes between them with rate
$\Lambda^A_{i\to f}$.
Straightforward algebra provides
\begin{equation}
\frac{1}{t_\mathrm{drain}^A}=
\Lambda^A_{i\to f}\frac{N_p}{N_i}+ \Lambda^A_{i\to f}e^{(E_f-E_i)/k_BT}\frac{N_p}{N_f}
\label{res:ratedescrip}
\end{equation}
where the backward process satisfying detailed balance
was also taken into account.
With $E_f-E_i=\SI{13.5}{meV}$, Fermi's golden rule provides
$\Lambda^A_{i\to f}= \SI{1.45}{ps^{-1}}$, which yields a drain time of
$t^A_\mathrm{drain}=\SI{0.46}{ps}$.
This agrees reasonably well with the fitted time scale above. The
discrepancy of \SI{0.08}{ps} may be
attributed to the thermalisation being on a time scale similar to the beating
between the different product states,
which is not included in the rate description of \Eref{res:ratedescrip}.

% Second initial state
Proceeding to the second initial state, $\ket{B}$, see Figure~\ref{fig:therm}(d),
we find that the fitted time-scale of draining of the hot-carrier levels 20-21 is
$\tau_\mathrm{drain}^B=\SI{26}{ps}$, and levels 6-9 and 10-11 are filled
in the time scales $\tau^B_\mathrm{fill1}=\SI{20}{ps}$ and
$\tau^B_\mathrm{fill2}=\SI{55}{ps}$ respectively.
In contrast to state  $\ket{A}$, level 4 is not occupied in
$\ket{B}$. Consequently, the observed draining and filling
time scales are much longer, as the
most relevant Coulomb scattering via
$V_{8,9,21,4}$ for $\ket{A}$ is not possible for state  $\ket{B}$.
Related decay paths to the states $|2,3,4,9\rangle$ or $|0,1,6,21\rangle$,
are provided by the elements $V_{4,9,21,0}$ or $V_{1,6,2,3}$, respectively.
The energy detuning of these are $\Delta E = \SI{-54}{meV}$ and
$\Delta E = \SI{43}{meV}$ (including charging energy), respectively,
which surpass the energy cut-off $D_\textrm{InP}=\SI{37}{meV}$.
Hence, these transitions are not individually accessible.
Yet, by combining both channels in a second order transition we obtain a total
detuning of $\Delta E = \SI{-14}{meV}$, well within the allowed range.
Other decay paths carry the same pattern of being excluded separately yet
allowed in combination due to the detuning considerations.
Hence, from our model assumptions, first order transitions are forbidden in this
thermalisation process, yet second order and higher combinations of these are
allowed.

Similarly to Sec.~\ref{SecBeating}, the second order transition rate from the
initial state $\ket{i}$ to a final state $\ket{f}$ over states $\ket{m}$
is calculated according to second order perturbation theory
\begin{eqnarray}
\Lambda^\mathrm{B}_{i\rightarrow f}
=\sum_{\ket{m}}\left|\frac{\bra{f}\hat{H}\ket{m}\bra{m}\hat{H}\ket{i}}
{E_i-E_m}\right|^2\frac{2\pi}{\hbar}\delta_{{\widetilde\Gamma}_\mathrm{deph}}(E_f-E_i).
\label{eq:incoherent}
\end{eqnarray}
Due to the $E_i - E_m$ term in the numerator above we cannot in general use
the Boltzmann term $e^{(E_f-E_i)/k_BT}$ to convert between the rates for
opposite directions.

Following the same procedure as for system A we consider a subset of the
Hilbert space with multiplicities $N_i=2$ for the initial set of states
with one electron in levels 20-21 and $N_f=2$ for the set of final states
$\{\ket{1,4,6,9},\ket{0,5,7,8}\}$ which are also the spin mirror of each other.
There are $N_p = 4$ Auger processes connecting the two sets,
however, unlike for system $A$ the individual processes
vary in strength. The full rate is thus
\begin{equation}
\frac{1}{t_\mathrm{drain}^B} = \sum_{i,f}
\left(\frac{\Lambda^B_{i\rightarrow f}}{N_i}
+\frac{\Lambda^B_{f\rightarrow i}}{N_f}\right)
\end{equation}
Summing over all possible transition channels we obtain the drain time
$t_\mathrm{drain}^B = \SI{43}{ps}$.
This is larger than the times observed on Fig.~\ref{fig:therm}(d).
This discrepancy may be attributed to the presence of four
other states with two electrons in levels 6-9 which compete for the
occupation from the initial states. The transition rates for these states and
our subset have the same order of magnitude and thus contribute to draining the
initial state on a relevant time scale. Hence, our subsystem reaches the
thermalised distribution quicker than our estimate which considers
the subsystem in isolation.

Note that \Eref{eq:incoherent} assumes an incoherent addition
of the different contributions to the same final state.
If we rather employ a coherent addition of the rates,
\begin{eqnarray*}
\Lambda^\mathrm{coh.}_{i\rightarrow f}
=\left|\sum_{\ket{m}}\frac{\bra{f}\hat{H}\ket{m}\bra{m}\hat{H}\ket{i}}
{E_i-E_m}\right|^2\frac{2\pi}{\hbar}\delta_{{\widetilde\Gamma}_\mathrm{deph}}(E_f-E_i)
\end{eqnarray*}
one obtains a drain time of \SI{498}{ps} which deviates by
an order of magnitude from the observed draining time.
As the dephasing time scale is
$\tau_\mathrm{deph}\sim\hbar/\Gamma_\mathrm{deph}=\SI{0.1}{ps}$ the observed
thermalisation is much slower than the dephasing making coherent addition
inapplicable. Additionally, coherent transitions are derived under the
assumption of unitary time evolution which is not the case for for the
Lindblad master equation, further explaining the necessity of incoherent addition.

The absence of first-order Coulomb scattering
in the second initial state results in  slow  thermalisation.
This makes us regard it as a worst-case scenario, with respect to quantum efficiency,
among possible initial states of the system.
As we will see in the subsequent section this renders the quantum efficiency
from the second initial state more difficult to augment above unity.

\subsection{Extraction through single barrier}
\label{sec:extract}
Above we found a difference of almost two orders of magnitude in the
thermalisation time scales between the two initial states under consideration.
We now include extraction operators from Eq.~(\ref{eq:meth:jump:extr})
in the Lindbladian time evolution in order to investigate the impact of
thermalisation on the extraction processes.

\begin{figure*}
\centering
\subfloat{\includegraphics[width=0.45\linewidth]{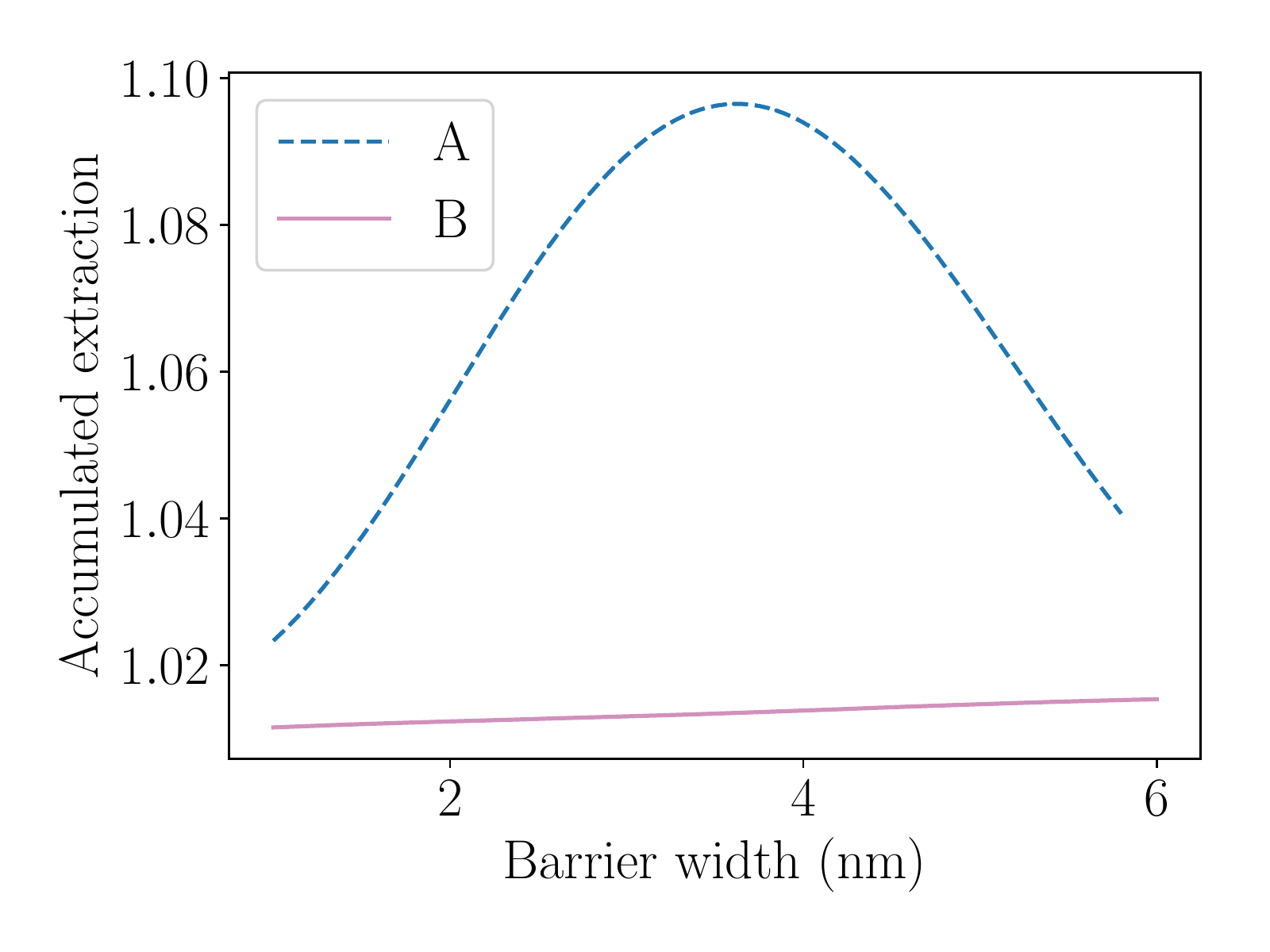}}
\hfill
\subfloat{\includegraphics[width=0.45\textwidth]{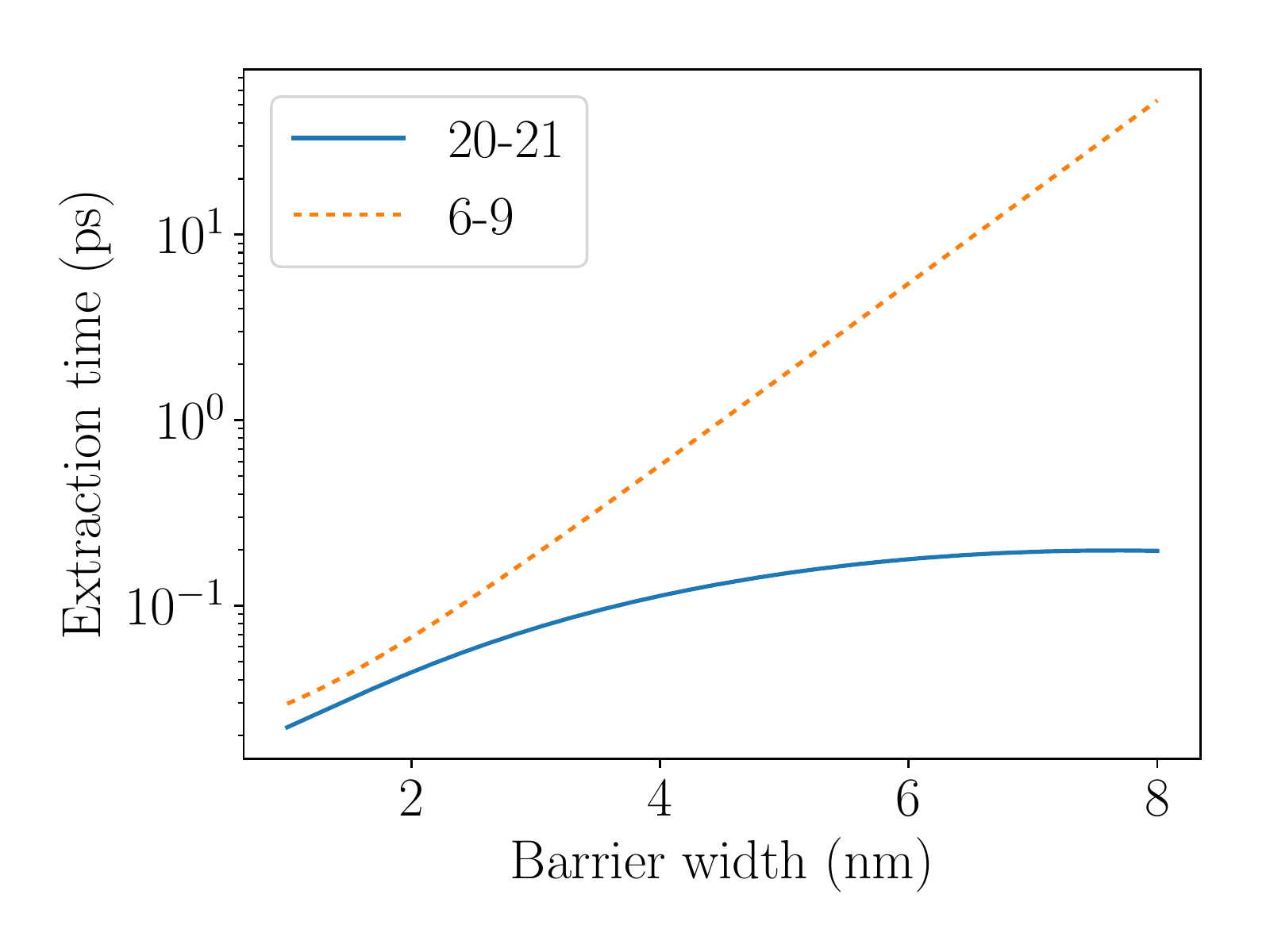}}
\caption{
Left: comparison of extraction from the QD through a single barrier potential
in the considered initial states $\ket{A}$ (dashed blue) and $\ket{B}$
(solid violet) after \SI{100}{ps}.
Right: extraction times for selected levels {in $\ket{A}$ and $\ket{B}$}.}
\label{fig:extract}
\end{figure*}

We want to investigate the effect of thermalisation whereby the
hot-carrier shares its energy with the three low-energy
electrons. Hence, we seek to exclude possible extraction contributions
stemming solely from the interaction between these three.  A
discernible Auger process among the three lower electrons is the decay
of the particle in level 3 into 1 which excites the other electron
from state 2 to a virtual state of \SI{489}{\milli\e\volt}.  To
preclude this contribution the potential offset in the nanowire is set
to ${\varrho=\SI{500}{\milli\e\volt}}$. Thus, no single
electron from the initial states $\ket{A}$, $\ket{B}$ can escape from
the QD unless the excited electron in level 21 participates.

Figure~\ref{fig:extract}(left) shows the accumulated particle
extraction for different barrier widths for both initial conditions.
Considering the initial state $\ket{A}$, dashed blue,
we find that the accumulated extraction has an optimum at a
barrier width of \SI{3.6}{nm}
with an electron extraction more than 0.09 above unity.
In order to understand the position of the optimum we consider the extraction
times of the selected orbitals, see Figure~\ref{fig:extract}(right).
At a barrier width around \SI{1}{nm} the extraction time is around
\SI{0.02}{\pico\second} for the upper levels 20-21.
This is three orders of magnitude smaller than the system
thermalisation time and thus the hot-carrier does not have sufficient time to
share a significant amount of its energy with the other electrons
before it is extracted. Hence, we have only a small increase of
electron extraction above unity.

As the barrier width is increased the extraction time of the high energy
electron, solid blue curve, increases. From the WKB approximation
the time increases approximately exponentially \cite{GriffithsBook2005},
\begin{equation}
\tau\sim T(E_e)^{-1} \sim \rme^{\kappa\Delta},
\quad \kappa = \sqrt{2{m_\mathrm{InP}}(V_0-E_e)}.
\end{equation}
{with parameters defined in sec. \ref{sec:meth:geom} and \ref{sec:meth:perlind}.}
This enables the hot-carrier to share more of its
energy with the other electrons, thus increasing the accumulated extraction.
However, from the WKB approximation we also see that the extraction time
is essentially dependent on the { decay parameter $\kappa$},
which is smaller for electrons with higher energy.
{Thus the extraction time for the levels 6-9, dashed orange,
increase quicker than that of the hot-carrier,}
which hinders the efficient
extraction from these states before the carrier is extracted from the QD.
If the levels 6-9 are emptied much slower than levels 20-21 the rate equations
from the previous subsection show that occupation flows back into the
hot-carrier levels, reducing the quantum efficiency.
These two opposing concerns appear to balance out just below a barrier width of
\SI{4}{nm}.

We now turn to the initial state $\ket{B}$, the total extraction after
\SI{100}{ps} is provided in Figure~\ref{fig:extract}(left).  We find that
the extraction never yields more than 0.02 above unity.  According to
Figure~\ref{fig:extract}(right) the extraction time of the hot-carrier
levels 20-21 never exceeds \SI{1}{ps} within the barrier widths of
interest, which is an order of magnitude faster than the
thermalisation time for this initial state, see the previous
subsection. While our model provides larger extraction
rates for barrier widths exceeding \SI{6}{nm} and longer collection times,
such results should not be taken too seriously. For time scales
approaching 100 ps, further interaction processes, not included here,
become of relevance for the thermalisation and will eventually limit
the extraction. Examples could be multiple-phonon interaction or
remote scattering with plasmons in gate contacts.

\subsection{Extraction through double barrier}
\label{sec:dbl}
\begin{figure*}
\centering
\includegraphics[width=0.8\linewidth]{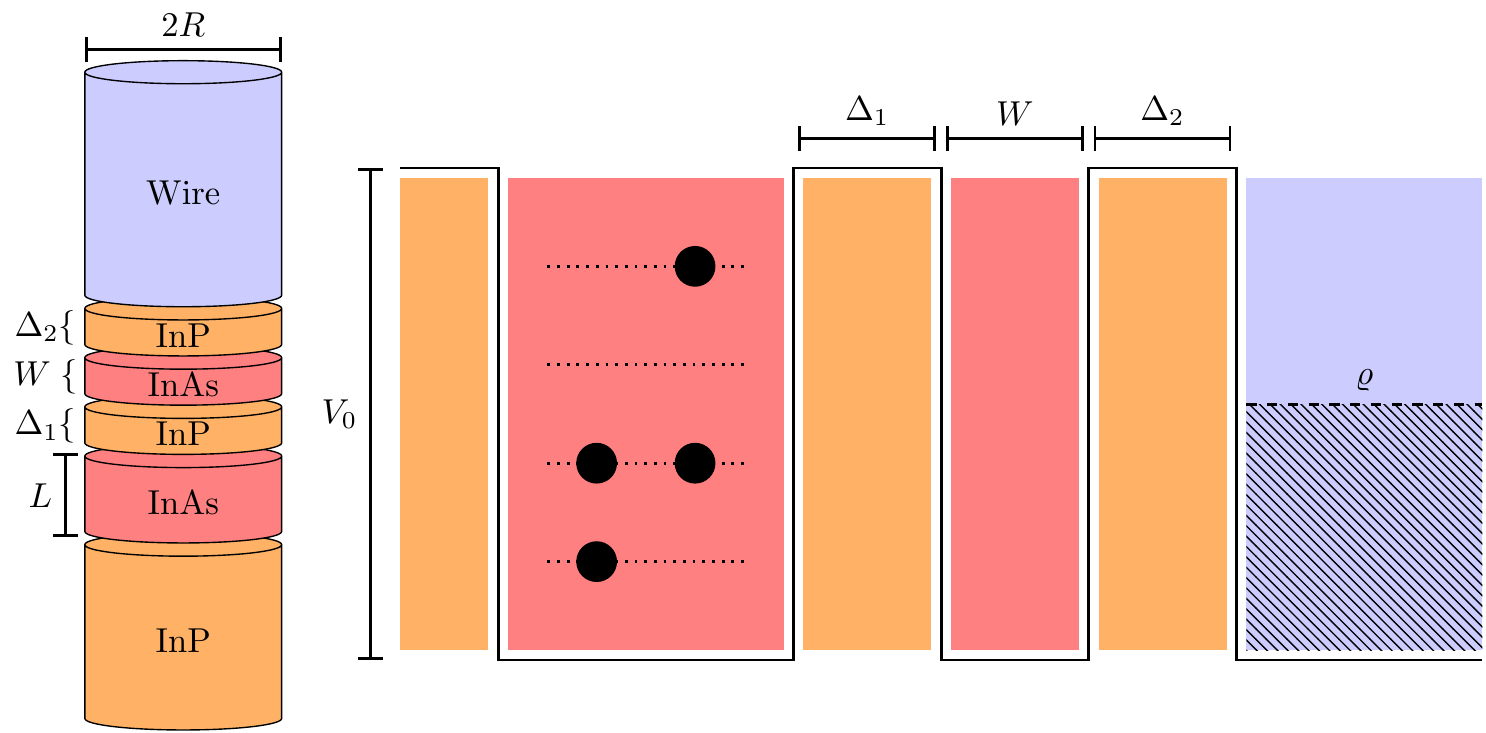}
\caption{Geometry and energy schematics of the nanowire QD system with double
barrier heterostructure. The QD is assumed to be a cylindrical piece of
InAs with height $L$ and radius $R$. To the one side it is connected to a thick
barrier of InP and to the other a superlattice structure of cylindrical
segments; an InP barrier of width $\Delta_1$, InAs well region
of width $W$ and then another InP barrier of width $\Delta_2$, before it
reaches the nanowire.}
\label{fig:dbl_schema}
\end{figure*}

We have identified the slow thermalisation of the initial state $\ket{B}$
to impair the total extraction through a single barrier due
to significant leaking of the hot-carrier.
Now we tailor the extraction region to improve the extraction from the levels
6-11, while disfavouring extraction from the initially occupied state 21.
This can be achieved by creating a double barrier region and adjusting
the widths according to resonant tunnelling for levels 6-11,
see Figure~\ref{fig:dbl_schema}.

\begin{figure*}
\centering
\subfloat{\includegraphics[width=0.45\linewidth]{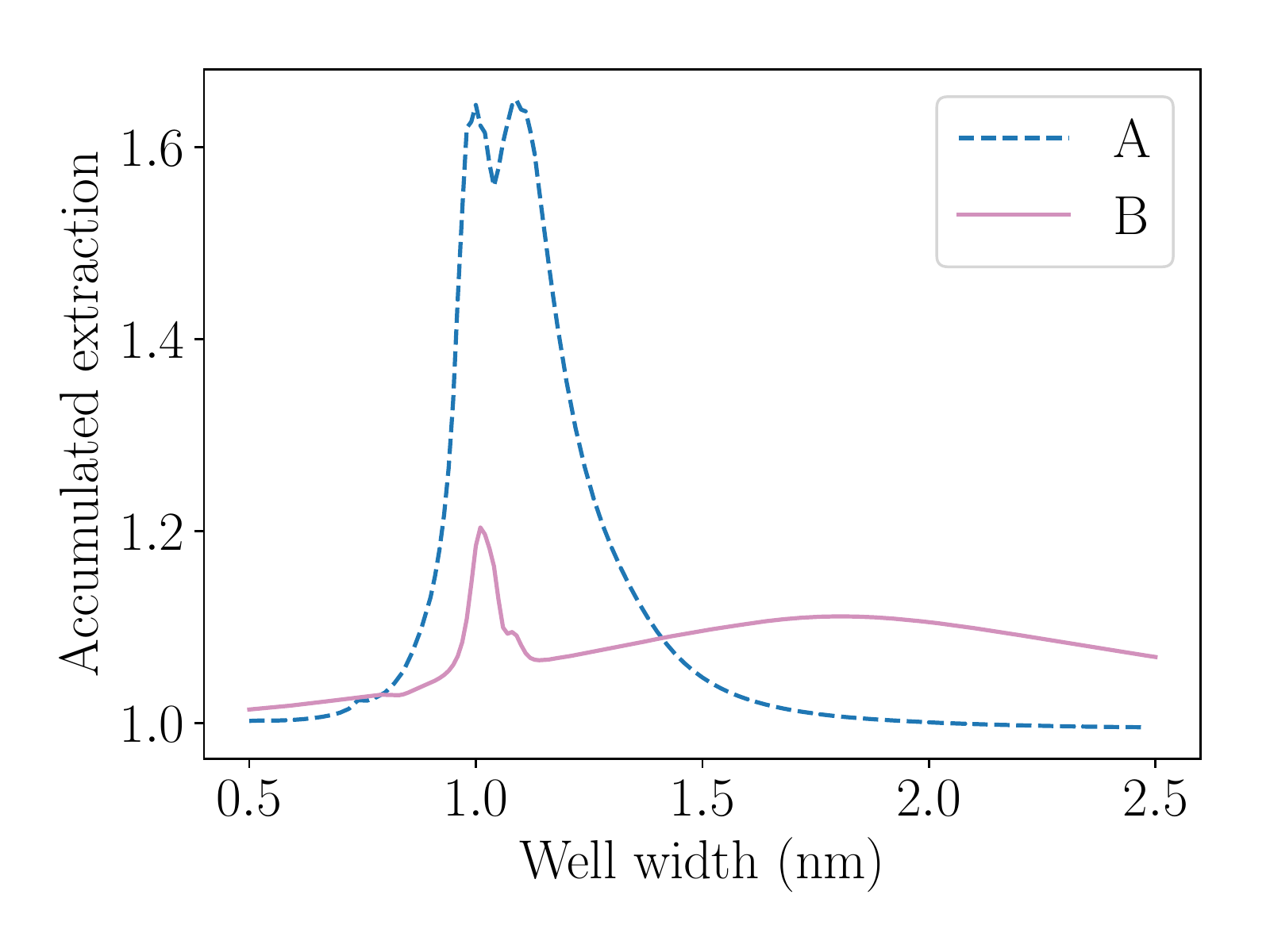}}
\hfill
\subfloat{\includegraphics[width=0.45\linewidth]{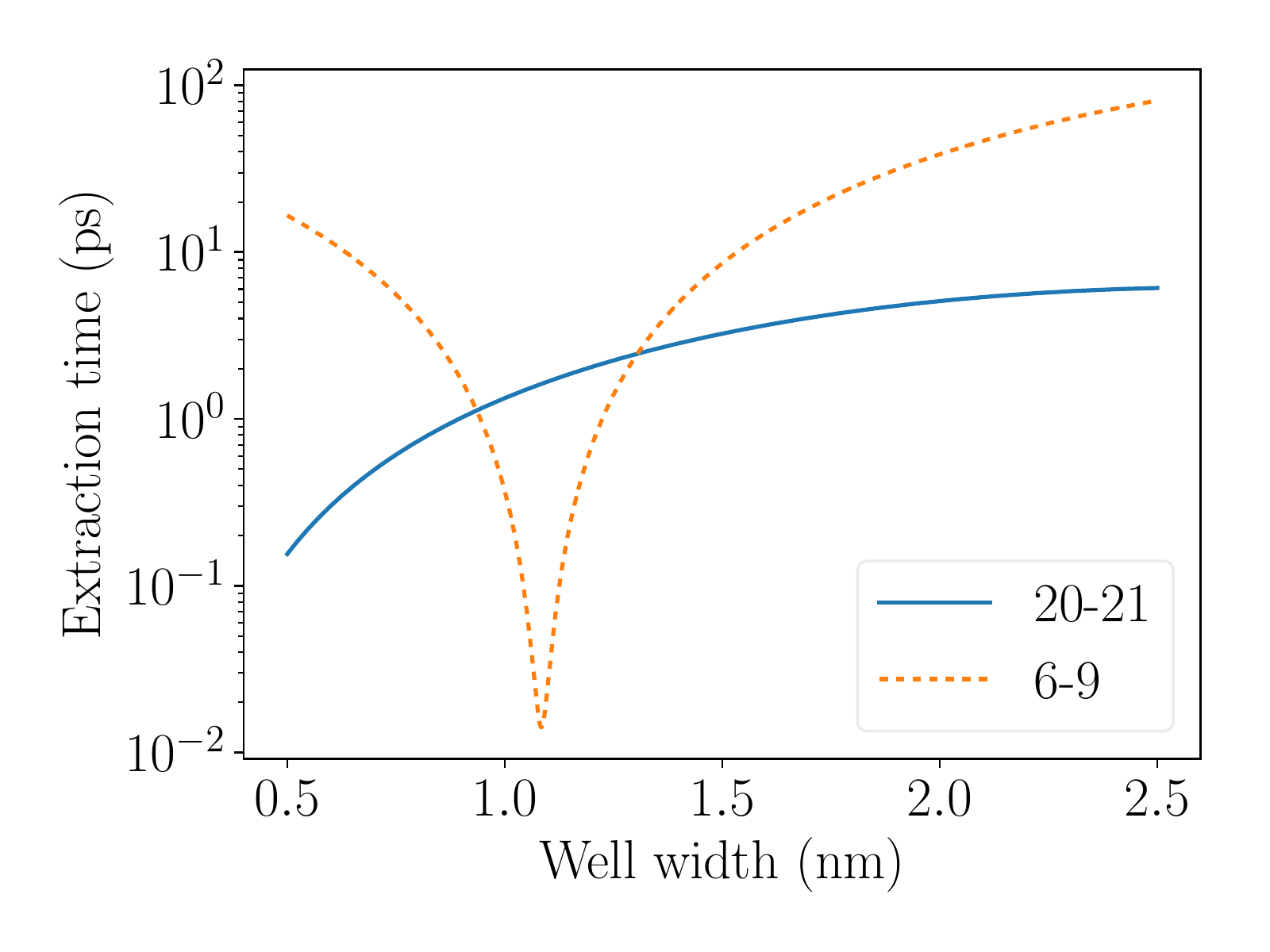}}
\caption{Extraction through a double barrier heterostructure from QD into
nanowire with barrier widths $\Delta_1=\Delta_2=\SI{4}{nm}$ and potential
offset in the nanowire $\varrho=\SI{500}{meV}$.
Left: Extraction after \SI{100}{\pico\second} for varying well widths.
Right: Extraction times {in $\ket{A}$ and $\ket{B}$} for selected levels for different well widths.}
\label{fig:dbl}
\end{figure*}

We let both barrier widths be $\Delta_1 = \Delta_2 = \SI{4}{\nano\meter}$
and vary the length of the of the well, $W$,
between the two barriers, see Figure~\ref{fig:dbl} (left).
The initial state is $\ket{B}$ and we
simulate the initial \SI{100}{\pico\second}.
From the accumulated extraction we see, for widths between 1 and
\SI{2}{\nano\meter}, a considerable increase in extracted electrons compared
to the single barrier extraction, cf. Figure~\ref{fig:extract}(b), with the
tallest peak at $W=\SI{1}{\nano\meter}$ where the extraction is
1.2 electrons. If we consider the extraction times for selected levels,
Figure~\ref{fig:dbl} (right), we find that the extraction time of the high
energy levels, solid blue curve, is monotonically increasing; above a width of
\SI{1}{\nano\meter} it is larger than \SI{2}{\pico\second} and thus within an
order of magnitude of the thermalisation time.
Additionally, the extraction times of the orbitals close to
\SI{500}{\milli\e\volt}, dashed green and dotted orange, are comparable to or
smaller than the thermalisation time in the interval between 1 and
\SI{1.7}{\nano\meter}. Below \SI{1}{\nano\meter} the
extraction time of the hot-carrier is too small to allow significant
redistribution of the energy while above \SI{2}{\nano\meter} the extraction
times of levels 6-11 are too large to allow significant extraction of
these states.
The pronounced peak in extraction around \SI{1}{\nano\meter} well width is due
to the tunnelling resonances for level 6-9 with the ground level in
the well. With perfect transmission for equal barriers, the extraction rate
approaches $\approx \SI{75}{\pico\second}^{-1}$, which provides a level
broadening of \SI{50}{\milli\e\volt}. The combined dephasing and extraction
broadening at resonance increases the detuning width of $\delta_\Gamma(\Delta E)$
which enables a faster
relaxation of the hot-carrier compared to pure dephasing.
Thus, relaxation and extraction processes cannot be treated
separately, if extraction rates become faster than \SI{0.1}{ps}.
  
Importantly, while the quantum efficiency of the double barrier potential
is particularly improved at resonance widths close to \SI{1}{nm},
the overall accumulated extraction is improved for all widths between
\SI{1.5}{nm} and \SI{2}{nm}, more than 0.1 above unity.
Even off-resonance the double barrier potential is a useful geometric configuration.
Due to the above considerations we 
expect that quickly thermalisable states will
likewise exhibit a pronounced increase in
quantum efficiency if we apply a similar double barrier potential.
{From Fig~\ref{fig:dbl} (left) we find this to be the case for state $\ket{A}$
with a broad peak reaching extractions of 0.6 above unity.}

%%%%%%%%%%%%%%%%%%%%%%%%%%%%%%%%%%%%%%%%%%%%%%%%
%
%	CONCLUSION
%
%%%%%%%%%%%%%%%%%%%%%%%%%%%%%%%%%%%%%%%%%%%%%%%%
\section{Conclusions}
The kinetics of photo-excited carriers in a nanowire quantum dot
system was studied. We find, that both first and second-order terms in
the electron-electron interactions are relevant, which is fully taken
into account by our PERLind approach applying a basis of many-particle
eigenstates of the quantum dot system.  Phenomenological dephasing and
tunnelling can be directly included and we find a strong dependence on
details of the system, such as the particular excited state.
{In particular, we quantify the thermalization due to second
  order processes, where a variety of different paths need to be
  added incoherently. We also point out, that very
efficient extraction can enable further transition processes due to
broadening, which is fully taken into account by our approach.}

If first-order processes in the
electron-electron scattering are energetically disfavoured our model
provides rather long thermalisation times {as expected}.
However, even in this
situation an increased extraction of electrons can be achieved by
including a double-barrier extraction structure. Here the efficiency
is not too sensitive to the actual resonance energy, so that
fluctuations in the well width of 10 \% do not affect the efficiency
dramatically { and the effect should be observable. For quick
thermalisation, when first order processes are possible, even higher total
extraction up to 1.6 are found. The extraction above unity
means, that the excess energy of the excitation above the band gap
due to a high energy photon can be converted into extra current,
and constitutes an interesting mechanism for advancing the efficiency
of the solar cells \cite{RossJAP1982,KolodinskiAPL1993}.}

{It would be interesting to study these processes experimentally by specifically designed optical pulses to provide specific single highly
  excited states similar to our states $|A\rangle$ and $|B\rangle$ in such nanowire structures. These experiments can be directly modeled by our approach, if the optical field is explicitly taken into account extending earlier work\cite{DamtieJChemPhys2016}. In this context it is also interesting to take into account
    non-ideal nanowire shapes seen in real systems, which, however, requires
    a substantially larger numerical effort.}

\ack
We thank the Knut and Alice Wallenberg foundation and NanoLund
for financial support.

\bibliographystyle{unsrt}

\end{document}